\documentclass{article}
\usepackage{spconf,amsmath,graphicx}
\usepackage{booktabs,amssymb}
\usepackage{dblfloatfix}
\usepackage{algorithm}
\usepackage{algpseudocode}
\usepackage{subcaption}
\usepackage{float}
\usepackage{placeins}
\usepackage{adjustbox}
\usepackage{enumitem}
\usepackage{hyperref}
\usepackage{xcolor}
\usepackage{soul}
\usepackage{nccmath}

\title{U-DAVI: Uncertainty-Aware Diffusion-Prior-Based Amortized Variational Inference for Image Reconstruction}
\name{Ayush Varshney, Katherine L. Bouman, Berthy T. Feng}
\address{Computing and Mathematical Sciences, California Institute of Technology}
\begin{document}
\ninept
\maketitle
\begin{abstract}
Ill-posed imaging inverse problems remain challenging due to the ambiguity in mapping degraded observations to clean images. Diffusion-based generative priors have recently shown promise, but typically rely on computationally intensive iterative sampling or per-instance optimization. Amortized variational inference frameworks address this inefficiency by learning a direct mapping from measurements to posteriors, enabling fast posterior sampling without requiring the optimization of a new posterior for every new set of measurements. However, they still struggle to reconstruct fine details and complex textures. To address this, we extend the amortized framework by injecting spatially adaptive perturbations to measurements during training, guided by uncertainty estimates, to emphasize learning in the most uncertain regions. Experiments on deblurring and super-resolution demonstrate that our method achieves superior or competitive performance to previous diffusion-based approaches, delivering more realistic reconstructions without the computational cost of iterative refinement.

\end{abstract}
\section{Introduction}
\label{sec:intro}

Ill-posed inverse problems in imaging, such as deblurring and super-resolution, require strong priors over natural images to regularize the solution space toward plausible results~\cite{bouman2016computational, ulyanov2018deep, menon2020pulse}. Diffusion models, first introduced as Denoising Diffusion Probabilistic Models (DDPM)~\cite{ho2020ddpm} and later generalized through stochastic differential equations~\cite{song2021score}, have emerged as principled, expressive, and robust generative priors~\cite{feng2023score}. A prominent line of work leverages these priors by conditioning the iterative reverse-sampling process on the measurement, which is typically achieved in one of two ways. The first approach involves guiding the sampling trajectory by incorporating measurement information at each step, often through likelihood gradients~\cite{chung2023diffusion, zhu2023denoising} or by optimizing a variational objective~\cite{mardani2024a}. The second approach uses analytical operators to enforce data consistency at each step, for instance through matrix decomposition~\cite{kawar2022denoising}, null-space projection~\cite{wang2023zero}, or pseudoinverse calculation~\cite{song2023pseudoinverse}. Although these techniques achieve high-fidelity results, their reliance on hundreds of function evaluations makes them computationally prohibitive for many applications.

Amortized inference offers a path to efficiency by training a single neural network to directly map a measurement to a posterior sample in one forward pass. The DAVI framework~\cite{lee2024davi} successfully implements this strategy for diffusion-based reconstruction, using a teacher--student score-matching objective (with the pre-trained diffusion prior as the teacher) to train an implicit generator that rivals iterative methods at a fraction of the cost. Unfortunately, this amortization comes with a well-known trade-off between inference speed and precision of the posterior approximation~\cite{cremer2018inference}. This trade-off appears in DAVI's results as reconstructions that are often inaccurate in ambiguous regions with fine textures and edges.

Recent work in diffusion-based iterative super-resolution has shown that modulating the noise level in specific regions based on an uncertainty estimate can improve the reconstruction of difficult details~\cite{zhang2025upsr}. However, applying such guidance in an amortized setting is challenging because it requires a lightweight uncertainty proxy. Conventional methods like deep ensembles~\cite{lakshminarayanan2017deep} or MC-dropout~\cite{gal2016dropout} are too expensive to run at every training step. A compelling alternative is inspired by self-ensembling techniques from semi-supervised learning, which use an exponential moving average (EMA) of past predictions as a stable training target~\cite{laine2017temporal, tarvainen2017meanteacher}. While their goal is to enforce consistency, we propose to repurpose this mechanism by reinterpreting the degree of temporal \emph{inconsistency} between the generator's current output and its historical EMA as an implicit, per-pixel measure of model uncertainty. This provides a useful guidance signal with negligible computational overhead.

In this work, we introduce Uncertainty-Aware DAVI (U\mbox{-}DAVI), a framework that integrates these ideas to enhance amortized variational inference without sacrificing its single-step efficiency. Our core contribution is an uncertainty-guided training curriculum. We first estimate a pixelwise uncertainty map by measuring the inconsistencies between the generator's current output and a persistent reconstruction memory maintained as an EMA per training sample. We then use this map to apply spatially adaptive noise within DAVI's Perturbed Posterior Bridge (PPB) mechanism, injecting stronger perturbations into uncertain regions. This forces the model to focus its capacity on resolving ambiguity, creating a self-refining curriculum that improves fine-detail reconstruction. Crucially, this guidance is only applied during training, preserving the single-step inference of the original amortized model.

Our contributions are summarized as follows:
\begin{enumerate}
    \item A lightweight, training-time uncertainty estimation mechanism based on temporal inconsistency with persistent reconstruction memories.
    \item An uncertainty-guided training strategy that leverages spatially adaptive noise to create a dynamic curriculum that targets difficult details.
    \item An amortized inference framework that delivers higher reconstruction quality and stronger generalization compared to previous variational inference approaches, without sacrificing single-step efficiency.
\end{enumerate}

\section{Method}
\label{sec:method}

\begin{figure}[t]
  \centering
  \includegraphics[page=1,width=\columnwidth, trim=0 5pt 0 0,clip]{\detokenize{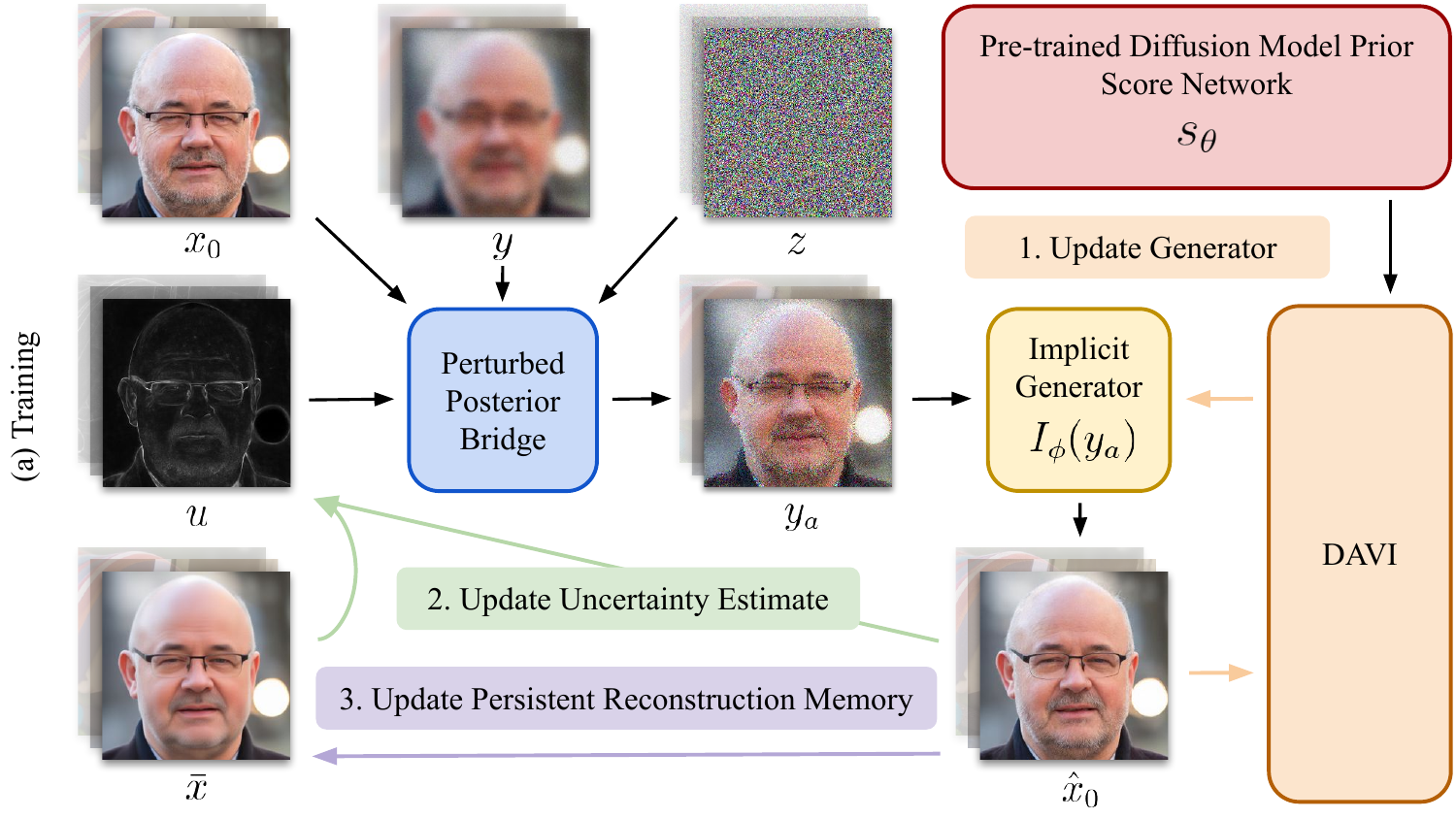}}
  \\[-0.6\baselineskip]
  \rule{\columnwidth}{0.45pt}\\[4pt]
  \includegraphics[page=2,width=\columnwidth,
  trim=0 285pt 0 0,clip]{\detokenize{results/U-DAVI_Figures.pdf}}

\caption{Uncertainty-Aware Diffusion-Prior-Based Amortized Variational Inference (U\mbox{-}DAVI) during (a) training and (b) inference. Notation: $x_0$ ground truth; $y$ noisy measurement; $z\!\sim\!\mathcal N(0,I)$; $u$ uncertainty estimate; $y_a$ uncertainty-aware bridge sample; $\hat x_0$ reconstruction; $\bar x$ persistent reconstruction memory. During training, the uncertainty map $u$ emphasizes ambiguous regions (e.g., around glasses). At inference, U\mbox{-}DAVI reconstructs $\hat x_0$ in a single pass and typically recovers fine details better than the original DAVI approach~\cite{lee2024davi}.\\[-7pt]}
  \label{fig:udavi_flowchart}
\end{figure}

We first review DAVI and then discuss our extensions for uncertainty estimation and guidance. Our framework, U\mbox{-}DAVI, operates as an uncertainty-aware fine-tuning phase applied to DAVI after it has converged (Figure~\ref{fig:udavi_flowchart}). This two-stage training strategy provides a stable baseline reconstruction, allowing our uncertainty mechanism to focus the generator's capacity specifically on resolving the most difficult and ambiguous regions of images.

\subsection{Background: The DAVI Framework}
\label{sec:DAVI_background}

We consider linear measurements $y = H x_0 + n$ with a known forward operator $H$ and measurement noise with a known distribution $n \sim \mathcal{N}(0, \sigma_y^2 I)$. DAVI aims to learn a single implicit generator, $I_{\phi}$, that approximates the true posterior $p(x_0|y)$. This generator is a neural network that maps a measurement $y$ and a random noise vector $z \sim \mathcal{N}(0, I)$ to a reconstruction sample $\hat{x}_0$, effectively sampling from the learned posterior $q_{\phi}(x_0|y)$.

\textbf{Variational Objective.}
The core objective is to optimize $\phi$ to minimize the Kullback--Leibler (KL) divergence from the learned posterior $q_{\phi}(x_0|y)$ to the true posterior $p(x_0|y)$:
\begin{fleqn}
\begin{equation}
\label{eq:main_objective}
\,\,\mathcal{L}(\phi) = \underbrace{-\mathbb{E}_{q_{\phi}(x_0|y)}[\log p(y|x_0)]}_{\text{Data Consistency}} + \underbrace{D_{KL}(q_{\phi}(x_0|y) \| p_{\theta}(x_0))}_{\text{Prior Regularization}}.
\end{equation}
\end{fleqn}

The first term is a data consistency loss, which reduces to an $\ell_2$ loss under Gaussian measurement noise:
\begin{equation}
\label{eq:consistency_loss}
\mathcal{L}_{C}(\phi) = \mathbb{E}_{\hat x_0 \sim q_{\phi}}\left[ \|y - H\hat{x}_0\|_2^2 \right].
\end{equation}

The second term enforces prior regularization with a pre-trained diffusion prior $p_{\theta}$ over clean images. This prior is defined implicitly by the forward noising process:
\begin{equation}
\label{eq:fwd_process}
x_t = \sqrt{\bar{\alpha}_t}\,x_0 + \sqrt{1-\bar{\alpha}_t}\,\epsilon,\quad \epsilon \sim \mathcal{N}(0, I),
\end{equation}
where $\bar{\alpha}_t = \prod_{i=1}^t (1-\beta_i)$ comes from the noise schedule $\{\beta_i\}_{i=1}^T$. Directly comparing $q_{\phi}(x_0|y)$ to $p_{\theta}(x_0)$ can be unstable when their supports do not overlap, so DAVI instead minimizes an Integral KL (IKL) divergence. IKL evaluates the mismatch after diffusing both distributions to intermediate states $x_t$, which effectively smooths them with Gaussian noise and guarantees overlap. Formally:
\begin{equation}
\label{eq:ikl_def}
\mathcal{L}_{IKL}(\phi) = \int_{t=0}^{T} w(t)\, D_{KL}(q_{\phi}(x_t|y) \| p_{\theta}(x_t))\,dt,
\end{equation}
with $w(t) = \sqrt{\bar{\alpha}_t}/\sqrt{1-\bar{\alpha}_t}$ placing more weight on low-noise timesteps where structural fidelity matters most. Its gradient is then approximated via teacher--student score matching:
\begin{equation}
\label{eq:ikl_loss_grad}
\nabla_{\phi} \mathcal{L}_{IKL} \propto \mathbb{E}_{\hat{x}_t, t} \left[ w(t)\,\underbrace{(s_{\psi}(\hat{x}_t, t) - s_{\theta}(\hat{x}_t, t))}_{\Delta s_{\psi, \theta}} \frac{\partial \hat{x}_t}{\partial \phi} \right],
\end{equation}
where $s_\theta$ is the fixed teacher score of the pre-trained diffusion model and $s_\psi$ is a learned student that estimates the score of the generator's evolving output distribution, $\nabla_{\hat{x}_t}\log q_\phi$. Separately, the student is trained by denoising score matching on $\hat{x}_t$ samples from $I_\phi$:
\begin{equation}
\label{eq:score_loss}
\mathcal{L}_{S}(\psi) = \mathbb{E}_{\hat{x}_t, t} \left[ \left\| s_{\psi}(\hat{x}_t, t) - \nabla_{\hat{x}_t} \log q_{\phi}(\hat{x}_t|y) \right\|_2^2 \right].
\end{equation}

Since both networks must be optimized jointly, DAVI alternates updates. At each training step, the generator produces $\hat{x}_0$, which is diffused to a noisy sample $\hat{x}_t$ at a random timestep $t \sim \mathcal{U}\{0,\dots,T\}$, ensuring supervision across the entire diffusion trajectory. Training proceeds by first updating $s_\psi$ using generator samples, followed by updating $I_\phi$ via $\nabla_{\phi}(\gamma\mathcal{L}_C + \mathcal{L}_{IKL})$, where $\gamma>0$ trades off data consistency and prior regularization. After training, $I_\phi$ can reconstruct any measurement in a single forward pass without any additional optimization, unlike iterative methods that require hundreds of refinement steps per reconstruction.

\textbf{Perturbed Posterior Bridge (PPB).}
To enhance generalization, DAVI is trained using samples from a stochastic bridge between clean images and raw measurements:
\begin{equation}
y_a = (1-\sigma_a)\,y + \sigma_a\,x_0 + h\,\bar{\sigma}_a\,z,
\quad z \sim \mathcal{N}(0, I).
\label{eq:ppb_davi}
\end{equation}
The sample's position along the bridge is defined by $a \in [0,1]$. The interpolation and stochasticity weights are based on the continuous-time formulation $\beta(\tau)$ of the discrete diffusion noise schedule:
\begin{equation}
\sigma_a = \frac{\int_a^1 \beta(\tau)\,d\tau}{\int_0^1 \beta(\tau)\,d\tau},
\quad
\bar{\sigma}_a = \sqrt{1-\exp\!\left(-\int_0^a \beta(\tau)\,d\tau\right)}.
\end{equation}
When $a=0$, the bridge sample $y_a$ is the clean image $x_0$, and when $a=1$, it becomes the measurement $y$ with a Gaussian perturbation scaled by a fixed hyperparameter $h$. As $a$ increases, $\sigma_a$ shifts weight toward the measurement while $\bar{\sigma}_a$ raises the stochasticity. The generator is trained to map these bridge inputs $y_a$ to reconstructions:
\begin{equation}
\label{eq:generator_pass}
\hat{x}_0 = I_{\phi}(y_a),
\end{equation}
so that during inference, when $a=1$, posterior samples can be drawn directly from $I_{\phi}(y + h z)$. 

Training with the PPB expands the model's input distribution around what it will see at test time while still incorporating signal from the clean image. This improves robustness to measurement variability and stabilizes variational optimization, resulting in higher-fidelity reconstructions.

\subsection{Uncertainty Estimation via Temporal Inconsistency}
\label{sec:temporal-inconsistency}
Our framework augments DAVI with a lightweight uncertainty estimate computed at each training iteration. For each training image $i$, we maintain a \emph{persistent reconstruction memory} $\bar{x}_i \in [0,1]^{H \times W \times 3}$, initialized to the pre-trained DAVI generator's output. The uncertainty map $u_i$ measures the pixelwise deviation between the current reconstruction and the persistent reconstruction memory.

Specifically, given the reconstruction $\hat{x}_{0,i} \in [-1,1]^{H \times W \times 3}$, we rescale to $[0,1]$ via $\hat{x}^{*}_{0,i} = (\hat{x}_{0,i}+1)/2$, and define per-pixel uncertainty as the normalized channel-wise $\ell_1$ distance from its memory:

\begin{equation}
\label{eq:uncertainty-map}
u_i(p) = \frac{\|\hat{x}^{*}_{0,i}(p)-\bar{x}_{i}(p)\|_{1}}{\max_{q} \|\hat{x}^{*}_{0,i}(q)-\bar{x}_{i}(q)\|_{1}},
\end{equation}
for all pixels $p$. Stable pixels yield $u_i(p) \approx 0$, while pixels whose reconstructions consistently fluctuate over time yield $u_i(p) \approx 1$.

After computing $u_i$, we update the memory for future iterations via an EMA with smoothing coefficient $\eta$, which determines the timescale of the memory:
\begin{equation}
\label{eq:ema-update}
\bar{x}_{i} \gets (1-\eta)\bar{x}_{i} + \eta\hat{x}^{*}_{0,i}.
\end{equation}
We relate $\eta$ to an effective memory window of $N$ updates using the standard formula $\eta=2/(N+1)$~\cite{brown1963smoothing}. 

Note that our uncertainty proxy uses the forward pass already required for training. In contrast, popular Bayesian approximations such as MC-dropout~\cite{gal2016dropout} and deep ensembles~\cite{lakshminarayanan2017deep} require $K$ stochastic passes or $M$ separate networks respectively. At our training scale, this overhead makes per-iteration guidance impractical. Our EMA-based temporal inconsistency approach provides a dense, per-pixel uncertainty estimate with only $O(1)$ additional compute every iteration and a small per-sample buffer.

\subsection{Uncertainty-Aware Perturbed Posterior Bridge}
\label{sec:uncertainty-guided-ppb}

The original PPB perturbs the entire measurement with i.i.d. Gaussian noise (Eq.~\ref{eq:ppb_davi}). We introduce a \emph{spatially adaptive perturbation} proportional to the uncertainty map $u_i$ from the previous iteration:
\begin{equation}
\label{eq:ppb_ours}
y_a = (1-\sigma_a)y
      + \sigma_a x_0
      + h\bar{\sigma}_a\,\bigl[\,z \odot (1+\lambda u_i)\bigr],
\end{equation}
where $z\sim\mathcal{N}(0, I)$, $\odot$ is element-wise multiplication, and $\lambda>0$ scales the uncertainty's effect on the perturbation.

Since the generator $I_{\phi}$ (a U-Net) tends to preserve spatial correspondence, amplified perturbations at high-uncertainty pixels mainly affect the corresponding regions of $\hat{x}_0$. After diffusing to time $t$, the resulting $\hat{x}_t$ continues to show greater variability in uncertain regions, which widens the difference $(s_\psi(\hat{x}_t,t) - s_\theta(\hat{x}_t,t))$. Through the IKL update in Eq.~\ref{eq:ikl_loss_grad}, this discrepancy generates stronger gradient signals for the generator in exactly those areas. Also, the amplified perturbations raise the residuals in the data consistency loss (Eq.~\ref{eq:consistency_loss}), reinforcing the focus on the same regions.

Since $u_i$ is recalculated at the end of every iteration, the uncertainty values evolve as training progresses. Pixels with stabilized predictions gradually move to lower values, while ambiguous regions shift to higher ones. As easier regions resolve, attention shifts toward the areas that remain difficult. This shifting focus creates a self-refining curriculum that directs learning effort where it is most needed, producing a generator that not only reliably reconstructs sharper details but also generalizes better across distribution shifts.

\subsection{Single-Step Inference}
\label{sec:inference}
The uncertainty estimation and guidance are exclusively training-time strategies designed to produce a more robust generator, so inference is still done with one forward pass:
\begin{equation}
\hat{x}_0 = I_{\phi}(y+hz), \quad z\sim\mathcal{N}(0, I).
\end{equation}

\section{Experiments}
\label{sec:experiments}

\subsection{Experimental Setup}
\textbf{Tasks.} Following DAVI~\cite{lee2024davi}, we evaluated two inverse problems: Gaussian deblurring and $4\times$ super-resolution. For deblurring, images were convolved with a $61\times 61$ Gaussian kernel (standard deviation $3.0$). For super-resolution, images were downsampled by a factor of $4$ using average pooling. In both tasks, we set the signal range to $[-1, 1]$ and added i.i.d.\ Gaussian measurement noise with standard deviation $\sigma_y = 0.05$. We trained and evaluated on the FFHQ 256$\times$256 dataset~\cite{karras2019stylegan}, following the DAVI protocol~\cite{lee2024davi} with 1{,}000 images for validation and 49{,}000 for training. For zero-shot generalization, we evaluated on the full 30{,}000-image CelebA-HQ $256\times256$ dataset~\cite{liu2015faceattributes,karras2018progressive}. We report PSNR for fidelity, Fr\'echet Inception Distance (FID)~\cite{heusel2017gans} for perceptual quality, and Number of Function Evaluations (NFE) for computational cost.

\textbf{Baselines.} Our primary baseline is DAVI~\cite{lee2024davi}, as our work directly extends this framework. We also compare against several iterative methods, including DDRM~\cite{kawar2022denoising}, DDNM$^+$~\cite{wang2023zero}, DPS~\cite{chung2023diffusion}, $\Pi$GDM~\cite{song2023pseudoinverse}, DiffPIR~\cite{zhu2023denoising}, and RED-Diff~\cite{mardani2024a}. For FFHQ, baseline results are taken directly from the DAVI paper to maintain a faithful and consistent comparison.

\textbf{Implementation Details.} All models were trained using AdamW with learning rate $1\times 10^{-4}$. By default, we fixed the PPB perturbation scale $h=0.1$, the bridge parameter $a \sim \mathrm{Beta}(3,1)$, and the persistent reconstruction memory window $N=8$. For Gaussian deblurring, we first trained DAVI for 42{,}000 iterations (batch size $8$, $T=400$, $\gamma=0.5$) to reproduce reported results, then continued training U\mbox{-}DAVI from DAVI's weights for 13{,}000 more iterations (batch size $16$, $\lambda=1.0$). For $4\times$ super-resolution, we first trained DAVI for 40{,}000 iterations (batch size $8$, $T=1000$, $\gamma=0.1$), then continued training U\mbox{-}DAVI from DAVI's weights for 4{,}000 more iterations (batch size $16$, $\lambda=0.5$).

% ===================== Figure 2 =====================
\begin{figure*}[t]
\centering
\setlength{\tabcolsep}{0pt}
\renewcommand{\arraystretch}{1.0}

% ---------- (a) Gaussian deblurring ----------
\begin{subfigure}[t!]{0.49\textwidth}
\centering
\begin{tabular}{@{}ccccc@{}}

\footnotesize Measurement & \footnotesize Ground Truth & \footnotesize RED\mbox{-}Diff & \footnotesize DAVI & \footnotesize U\mbox{-}DAVI (Ours) \\[2pt]

\includegraphics[width=\dimexpr\linewidth/5\relax]{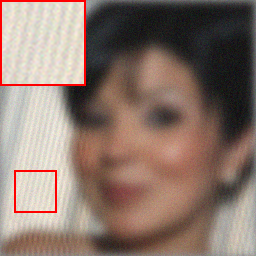} &
\includegraphics[width=\dimexpr\linewidth/5\relax]{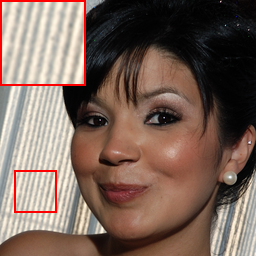} &
\includegraphics[width=\dimexpr\linewidth/5\relax]{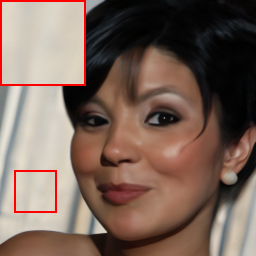} &
\includegraphics[width=\dimexpr\linewidth/5\relax]{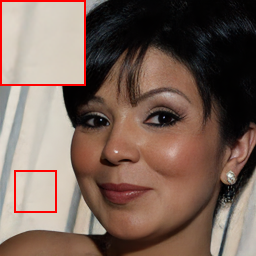} &
\includegraphics[width=\dimexpr\linewidth/5\relax]{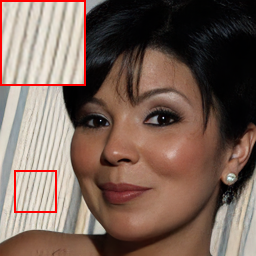} \\[-3pt]

\includegraphics[width=\dimexpr\linewidth/5\relax]{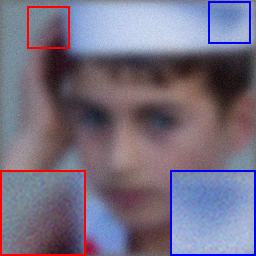} &
\includegraphics[width=\dimexpr\linewidth/5\relax]{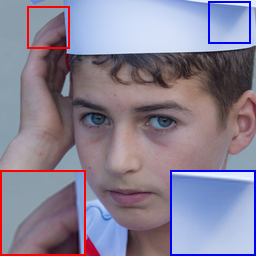} &
\includegraphics[width=\dimexpr\linewidth/5\relax]{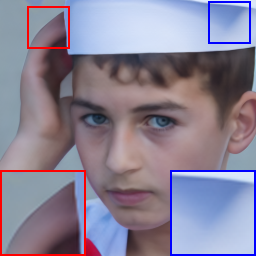} &
\includegraphics[width=\dimexpr\linewidth/5\relax]{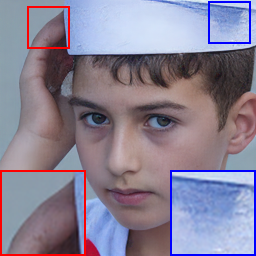} &
\includegraphics[width=\dimexpr\linewidth/5\relax]{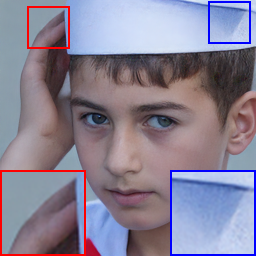} \\[-3pt]

\includegraphics[width=\dimexpr\linewidth/5\relax]{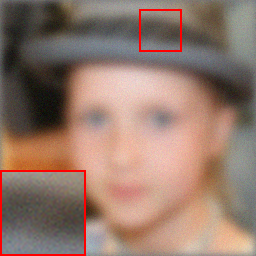} &
\includegraphics[width=\dimexpr\linewidth/5\relax]{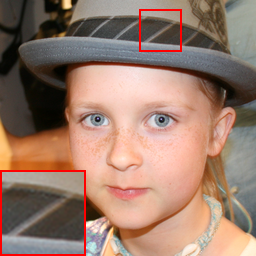} &
\includegraphics[width=\dimexpr\linewidth/5\relax]{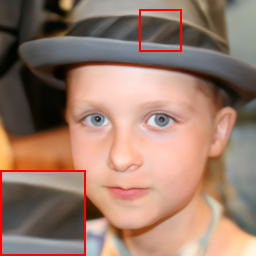} &
\includegraphics[width=\dimexpr\linewidth/5\relax]{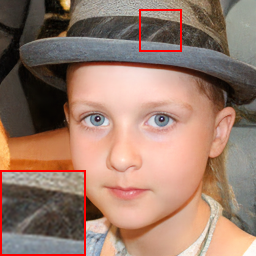} &
\includegraphics[width=\dimexpr\linewidth/5\relax]{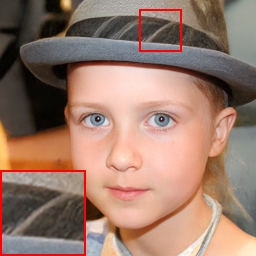}
\end{tabular}\\[-3pt]
\subcaption{Gaussian deblurring (61$\times$61 kernel, $\sigma=3.0$)}
\end{subfigure}
\hfill
% ---------- (b) 4x Super-resolution ----------
\begin{subfigure}[t!]{0.49\textwidth}
\centering
\begin{tabular}{@{}ccccc@{}}
\footnotesize Measurement & \footnotesize Ground Truth & \footnotesize RED\mbox{-}Diff & \footnotesize DAVI & \footnotesize U\mbox{-}DAVI (Ours) \\[2pt]

\includegraphics[width=\dimexpr\linewidth/5\relax]{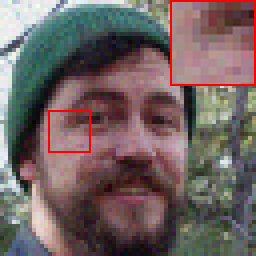} &
\includegraphics[width=\dimexpr\linewidth/5\relax]{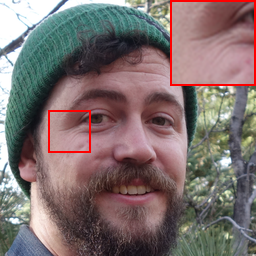} &
\includegraphics[width=\dimexpr\linewidth/5\relax]{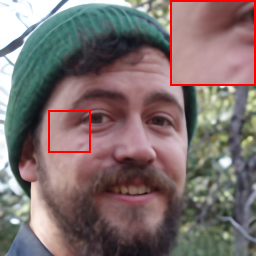} &
\includegraphics[width=\dimexpr\linewidth/5\relax]{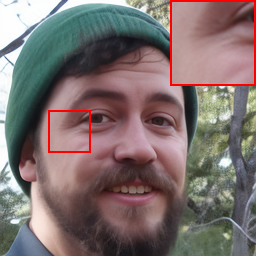} &
\includegraphics[width=\dimexpr\linewidth/5\relax]{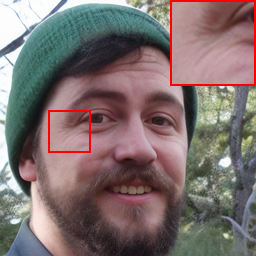} \\[-3pt]

\includegraphics[width=\dimexpr\linewidth/5\relax]{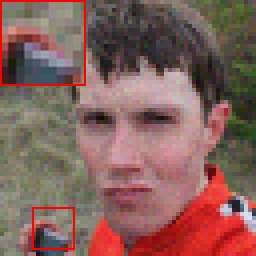} &
\includegraphics[width=\dimexpr\linewidth/5\relax]{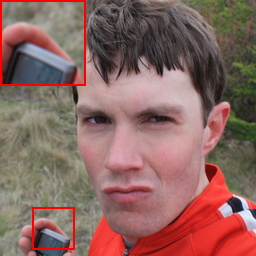} &
\includegraphics[width=\dimexpr\linewidth/5\relax]{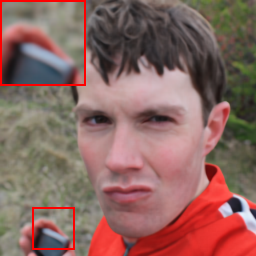} &
\includegraphics[width=\dimexpr\linewidth/5\relax]{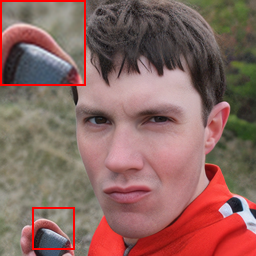} &
\includegraphics[width=\dimexpr\linewidth/5\relax]{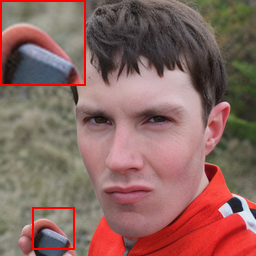} \\[-3pt]

\includegraphics[width=\dimexpr\linewidth/5\relax]{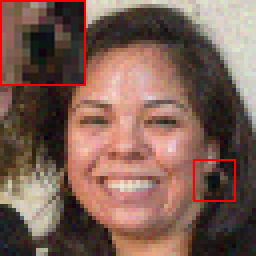} &
\includegraphics[width=\dimexpr\linewidth/5\relax]{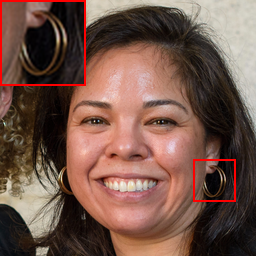} &
\includegraphics[width=\dimexpr\linewidth/5\relax]{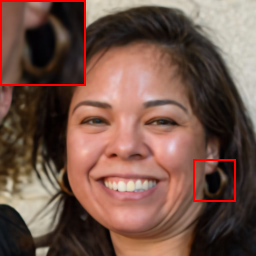} &
\includegraphics[width=\dimexpr\linewidth/5\relax]{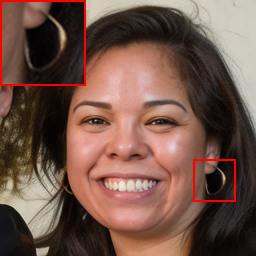} &
\includegraphics[width=\dimexpr\linewidth/5\relax]{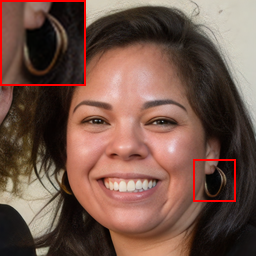}
\end{tabular}\\[-3pt]
\subcaption{$4\times$ Super-resolution (average pooling)}
\end{subfigure}

\caption{Qualitative results on FFHQ for (a) deblurring and (b) super-resolution. In both tasks, U\mbox{-}DAVI restores sharper details such as backgrounds, wrinkles, fingers, shadows, reflections, fabrics, and earrings compared to the iterative RED\mbox{-}Diff and amortized DAVI baselines.\\[-7pt]}

\label{fig:inverse_problems_comparison}
\end{figure*}
% =================== End Figure 2 =====================

\subsection{Discussion}
\label{sec:discussion}

\textbf{In-Distribution Performance.} 
On the FFHQ validation set (Table~\ref{tab:ffhq_tab}), U\mbox{-}DAVI consistently improved upon DAVI for both Gaussian deblurring and $4\times$ super-resolution. The most substantial gains were in FID, showing that our uncertainty-guided training enhances perceptual realism without sacrificing fidelity. Unlike iterative methods, which demand hundreds of NFEs, U\mbox{-}DAVI reaches competitive PSNR and superior FID in a single forward pass. Qualitative comparisons (Figure~\ref{fig:inverse_problems_comparison}) illustrate that U\mbox{-}DAVI reconstructed realistic material textures, facial wrinkles, and lighting details unlike RED-Diff and DAVI, resulting in sharper and more visually convincing images.

\textbf{Out-of-Distribution Generalization.} 
Zero-shot evaluation on CelebA\mbox{-}HQ (Table~\ref{tab:celebahq_tab}) tests robustness to distribution shift. Here, U\mbox{-}DAVI continued to surpass DAVI, with better PSNR and FID in both tasks. We omit iterative baselines to keep the comparison fair under zero\mbox{-}shot: freezing FFHQ\mbox{-}tuned settings yields under\mbox{-}tuned results on CelebA\mbox{-}HQ, while re\mbox{-}tuning would violate zero\mbox{-}shot and require extensive compute sweeps. These drawbacks also highlight the practical advantage of amortized inference for deployment.

% --- FFHQ (single-column) ---
\begin{table}[t]
  \centering
  \resizebox{\columnwidth}{!}{%
  \begin{tabular}{@{}l c cc cc@{}}
    \toprule
    \multicolumn{2}{c}{} &
      \multicolumn{2}{c}{Gaussian deblur} &
      \multicolumn{2}{c}{$4\times$ SR} \\[1pt]
    Method & NFE $\downarrow$ &
      PSNR $\uparrow$ & FID $\downarrow$ &
      PSNR $\uparrow$ & FID $\downarrow$ \\
    \midrule
    DDRM~\cite{kawar2022denoising}         & 20   & 26.26 & 56.82 & 28.09 & 48.19 \\
    DDNM$^+$~\cite{wang2023zero}           & 100  & 24.19 & 91.48 & 28.17 & 59.09 \\
    DPS~\cite{chung2023diffusion}          & 1000 & 21.88 & 34.47 & 25.55 & 36.29 \\
    $\Pi$GDM~\cite{song2023pseudoinverse}  & 100  & 23.05 & 52.72 & 27.73 & 49.86 \\
    DiffPIR~\cite{zhu2023denoising}        & 100  & 24.41 & 33.91 & 25.32 & 40.35 \\
    RED\mbox{-}Diff~\cite{mardani2024a}    & 1000 & \textbf{26.44} & 46.55 & 26.75 & 92.82 \\
    \midrule
    DAVI~\cite{lee2024davi}                & \textbf{1} & 25.46 & 29.80 & 28.23 & 23.73 \\
    U\mbox{-}DAVI (Ours)                   & \textbf{1} & 25.61 & \textbf{28.97} & \textbf{28.24} & \textbf{23.62} \\
    \bottomrule
  \end{tabular}%
  }
  \caption{Validation set results on FFHQ. U\mbox{-}DAVI achieves lower FID than DAVI on both tasks and slightly improves PSNR. Iterative baselines are included for context and generally show worse FID despite requiring orders of magnitude more compute during inference.\\[-7pt]}
  \label{tab:ffhq_tab}
\end{table}

\begin{figure}[t!]
    \centering
    \setlength{\tabcolsep}{3pt}

    \begin{tabular}{@{}c|c|c@{}}
        & FFHQ & CelebA\mbox{-}HQ (zero-shot) \\
        \hline
        % Row 1: Gaussian deblurring
        \raisebox{-0.4\height}{\rotatebox[origin=c]{90}{\parbox{1.6cm}{\centering Gaussian deblurring}}} &
        \raisebox{-0.4\height}{%
          \begin{tabular}[c]{@{}c@{\hspace{1pt}}c@{}}
            % ---- space right after begin{tabular} ----
            \multicolumn{2}{@{}c@{}}{\rule{0pt}{6pt}}\\[-7pt]
            \shortstack[c]{\includegraphics[width=0.2\linewidth]{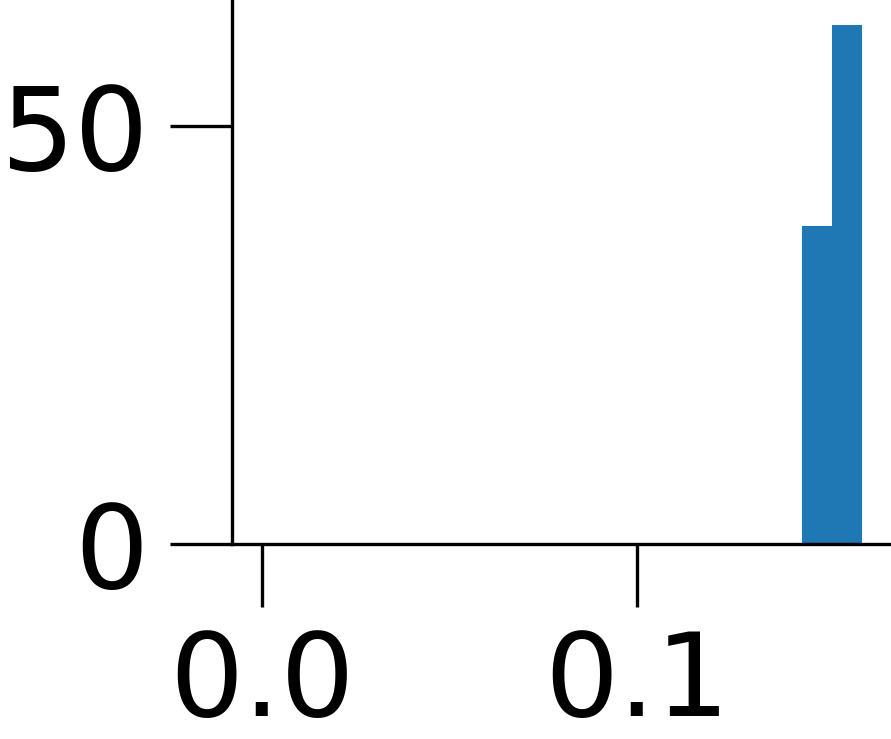} \\ \scriptsize $\Delta$PSNR} &
            \shortstack[c]{\includegraphics[width=0.2\linewidth]{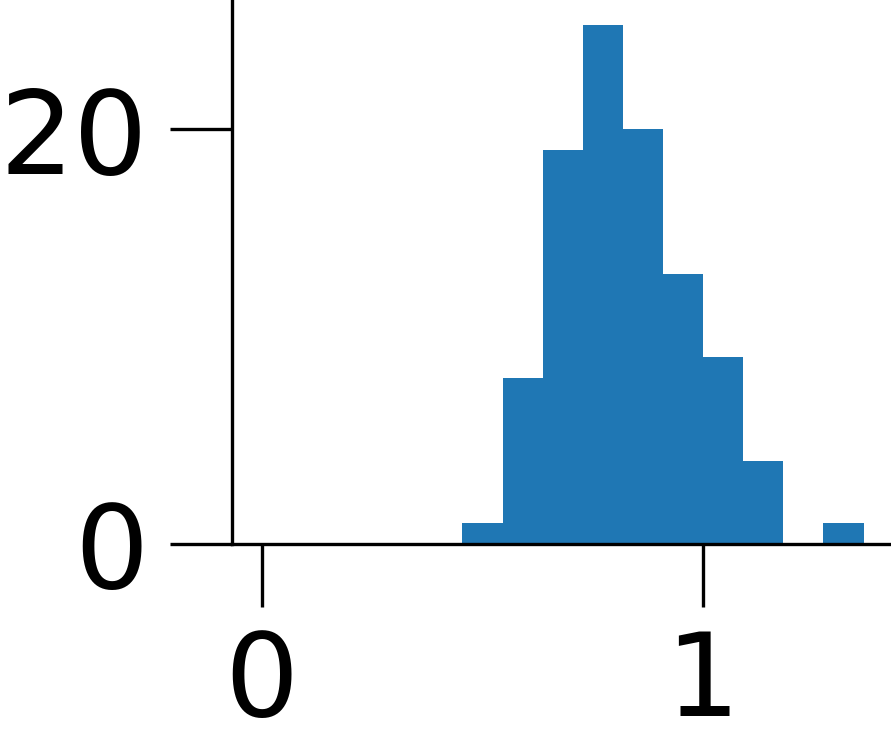} \\ \scriptsize $\Delta$FID}
          \end{tabular}%
        } &
        \raisebox{-0.4\height}{%
          \begin{tabular}[c]{@{}c@{\hspace{1pt}}c@{}}
            % ---- space right after begin{tabular} ----
            \multicolumn{2}{@{}c@{}}{\rule{0pt}{6pt}}\\[-7pt]
            \shortstack[c]{\includegraphics[width=0.2\linewidth]{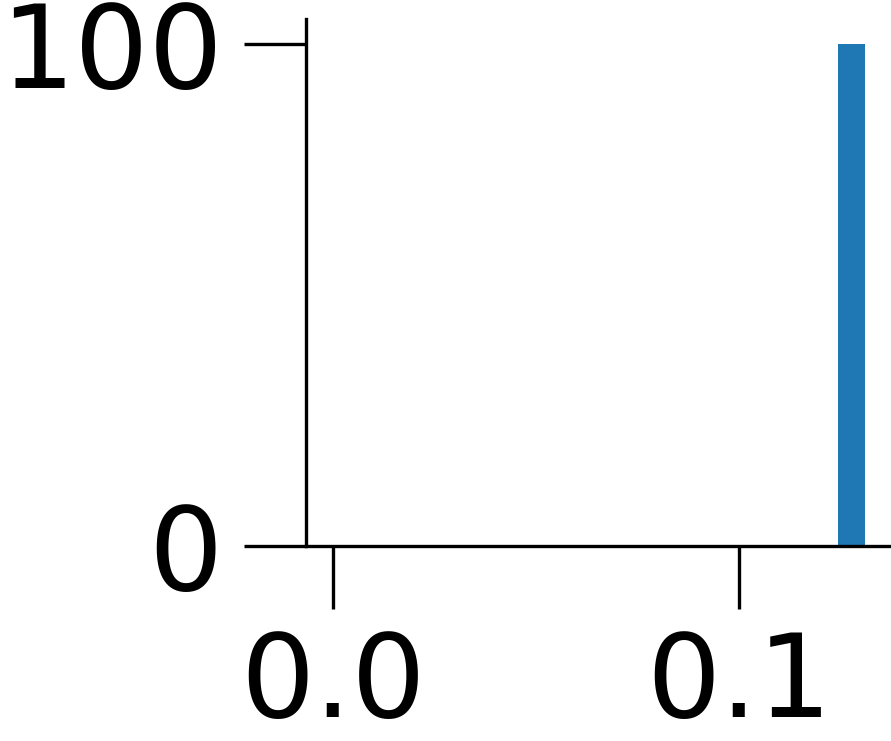} \\ \scriptsize $\Delta$PSNR} &
            \shortstack[c]{\includegraphics[width=0.2\linewidth]{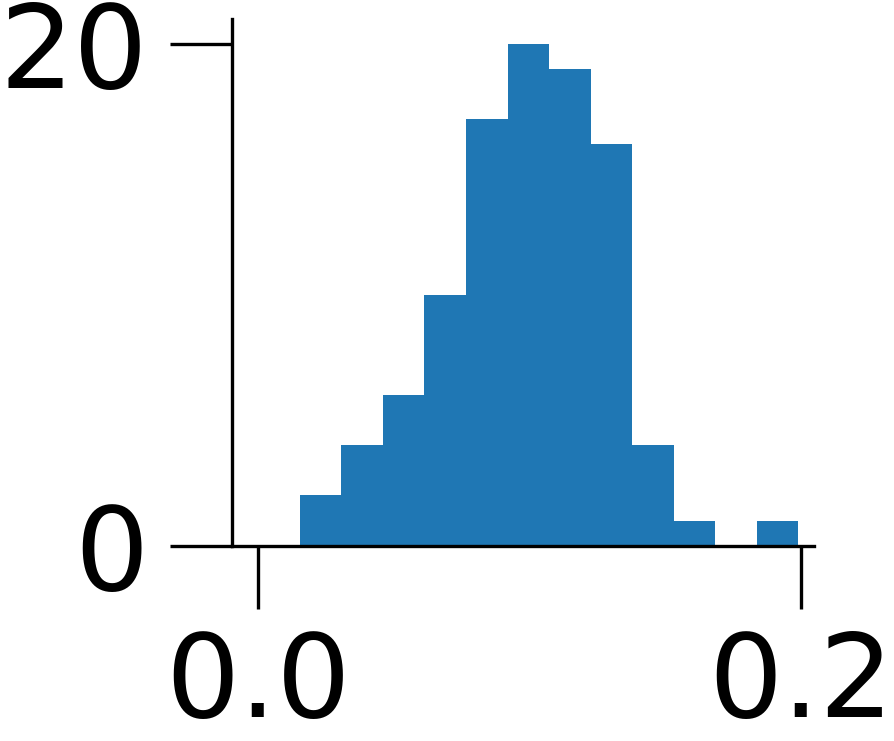} \\ \scriptsize $\Delta$FID}
          \end{tabular}%
        } \\
        \hline
        % Row 2: 4x Super-resolution
        \raisebox{-0.4\height}{\rotatebox[origin=c]{90}{\parbox{1.6cm}{\centering $4\times$ Super-resolution}}} &
        \raisebox{-0.4\height}{%
          \begin{tabular}[c]{@{}c@{\hspace{1pt}}c@{}}
            % ---- space right after begin{tabular} ----
            \multicolumn{2}{@{}c@{}}{\rule{0pt}{6pt}}\\[-7pt]
            \shortstack[c]{\includegraphics[width=0.2\linewidth]{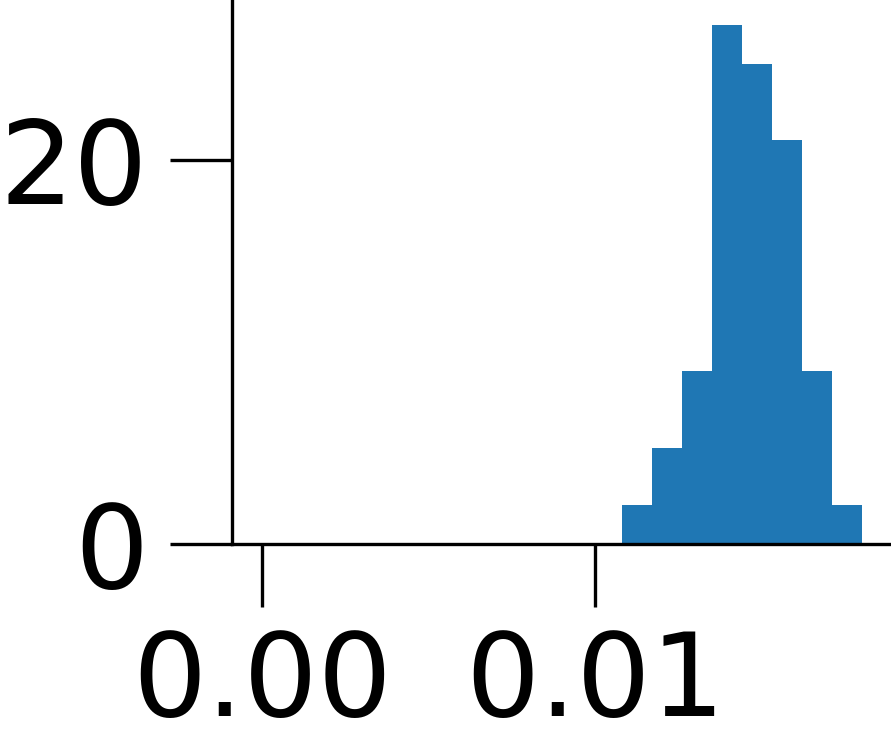} \\ \scriptsize $\Delta$PSNR} &
            \shortstack[c]{\includegraphics[width=0.2\linewidth]{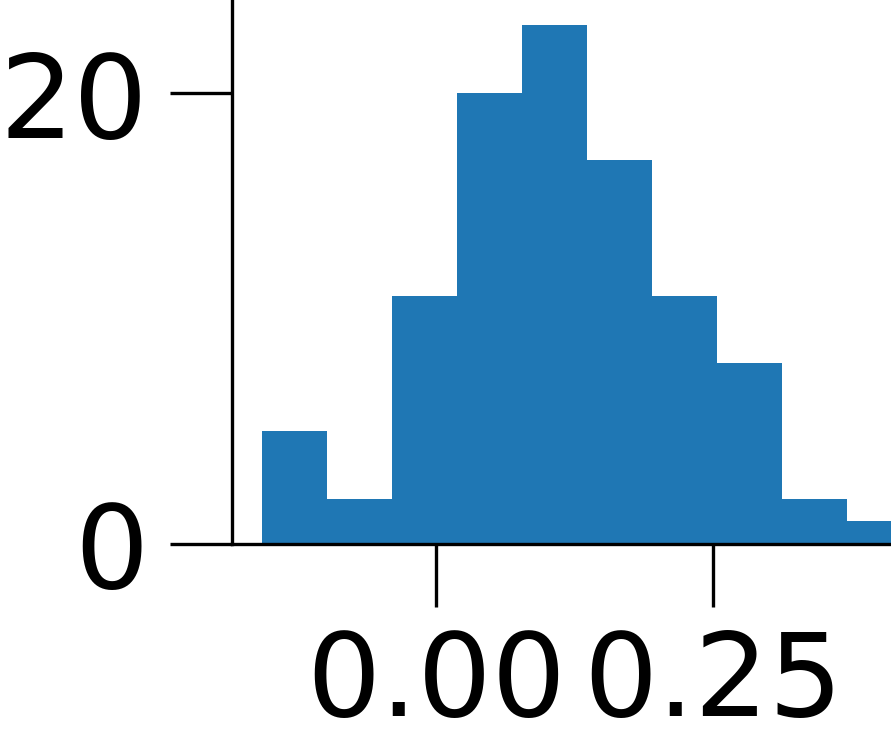} \\ \scriptsize $\Delta$FID}
          \end{tabular}%
        } &
        \raisebox{-0.4\height}{%
          \begin{tabular}[c]{@{}c@{\hspace{1pt}}c@{}}
            % ---- space right after begin{tabular} ----
            \multicolumn{2}{@{}c@{}}{\rule{0pt}{6pt}}\\[-7pt]
            \shortstack[c]{\includegraphics[width=0.2\linewidth]{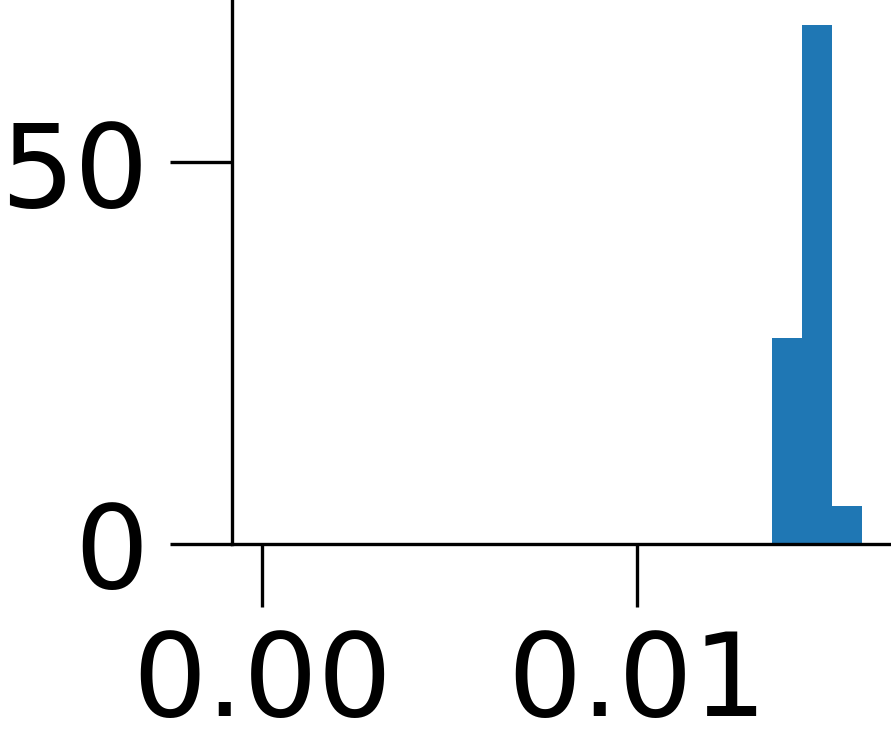} \\ \scriptsize $\Delta$PSNR} &
            \shortstack[c]{\includegraphics[width=0.2\linewidth]{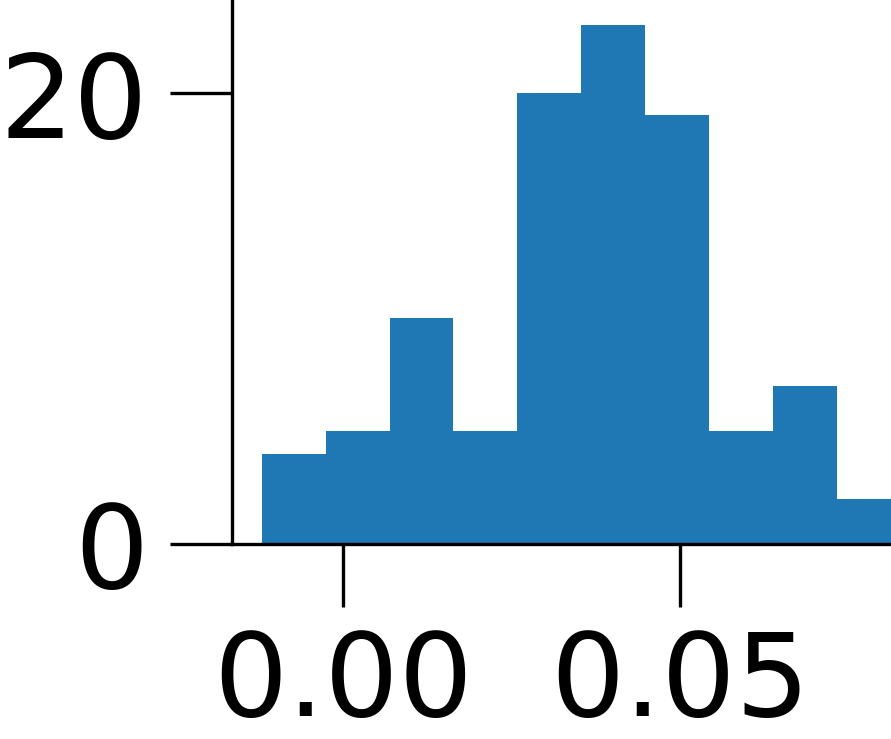} \\ \scriptsize $\Delta$FID}
          \end{tabular}%
        }
    \end{tabular}

    \caption{Histograms of U\mbox{-}DAVI's improvements on deblurring and super-resolution for FFHQ (1K images) and CelebA\mbox{-}HQ (30K images) across 100 inference seeds. We plot $\Delta\text{PSNR}=\text{PSNR}_{\text{U\mbox{-}DAVI}}-\text{PSNR}_{\text{DAVI}}$ and $\Delta\text{FID}=\text{FID}_{\text{DAVI}}-\text{FID}_{\text{U\mbox{-}DAVI}}$ such that positive values indicate U\mbox{-}DAVI outperforms DAVI (higher PSNR / lower FID).\\[-7pt]}
    \label{fig:histograms}
\end{figure}

% --- CelebA-HQ (single-column) ---
\begin{table}[t!]
  \centering
  \resizebox{\columnwidth}{!}{%
  \begin{tabular}{@{}l c cc cc@{}}
    \toprule
    \multicolumn{2}{c}{} &
      \multicolumn{2}{c}{Gaussian deblur} &
      \multicolumn{2}{c}{$4\times$ SR} \\[1pt]
    Method & NFE $\downarrow$ &
      PSNR $\uparrow$ & FID $\downarrow$ &
      PSNR $\uparrow$ & FID $\downarrow$ \\
    \midrule
    DAVI~\cite{lee2024davi}                & \textbf{1} & 26.22 & 26.43 & 29.23 & 25.81 \\
    U\mbox{-}DAVI (Ours)                   & \textbf{1} & \textbf{26.35} & \textbf{26.33} & \textbf{29.24} & \textbf{25.77} \\
    \bottomrule
  \end{tabular}%
  }
  \caption{Zero-shot results on entire CelebA\mbox{-}HQ. Without retraining, U\mbox{-}DAVI still delivers lower FID and higher PSNR than DAVI.\\[-7pt]}
  \label{tab:celebahq_tab}
\end{table}

\textbf{Ablation on Uncertainty Scaling.} The hyperparameter $\lambda$ controls the adaptive perturbation scale in U\mbox{-}DAVI's training (Eq.~{\ref{eq:ppb_ours}}). Higher $\lambda$ improved PSNR but degraded FID, while lower $\lambda$ showed the opposite trend. This was likely because slight uncertainty-guided perturbations stabilize predictions in ambiguous regions, whereas large perturbations blur signal and bias the model toward smoother outputs. We tested $\lambda\in\{0.5,1.0,2.0\}$ and selected $\lambda=1.0$ for deblurring and $\lambda=0.5$ for super-resolution as the best trade-offs.

\textbf{Ablation on Memory Window.} The hyperparameter $N$ determines how quickly persistent reconstruction memories adapt to changes in the generator's output every iteration (Eq.~{\ref{eq:ema-update}}). U\mbox{-}DAVI reached its peak performance earlier with $N=4$ and later with $N=16$ compared to when it used $N=8$. We found that this peak performance was very similar across all three settings and chose to report results using $N=8$. 

\textbf{Reliability of Results.} We repeated evaluations with 100 random seeds at inference time for both DAVI and U\mbox{-}DAVI across all combinations of dataset, task, and metric. Paired t-tests show that improvements are statistically significant ($p < 0.0001$) in all cases. Figure~{\ref{fig:histograms}} visualizes the distributions of per-seed improvements and shows they are clearly centered in the positive domain for both PSNR and FID, validating that our findings are not due to randomness.

\section{Conclusion}
\label{sec:conclusion}

In this paper, we introduced U\mbox{-}DAVI, an uncertainty-aware framework for solving imaging inverse problems. By augmenting the amortized variational inference of DAVI with an uncertainty estimation and guidance mechanism, our method learns to focus on the most challenging regions of an image during training. The key advantage of our approach is that it translates into a more robust model, which still performs inference in a single, efficient forward pass.

Our experiments across two image restoration tasks show that U\mbox{-}DAVI improves upon the strong DAVI baseline and many other diffusion-based methods, particularly in terms of perceptual quality. Furthermore, our zero-shot evaluation on an out-of-distribution dataset highlights the enhanced generalization capability of our model. By confronting and resolving its predictive uncertainty, U\mbox{-}DAVI learns to produce sharper, more faithful reconstructions, pushing the boundaries of what is possible with single-step, diffusion-based inverse problem solvers.

\section{Acknowledgments}
\label{sec:acknowledgments}
This work was sponsored by NSF Award 2048237, an Amazon AI4Science Discovery Award, OpenAI, a Sloan Research Fellowship, and a Pritzker Award. We also thank Christina Liu and Jeremy Budd for helpful discussions early on.

\FloatBarrier

\bibliographystyle{IEEEtran}
\bibliography{main}

\end{document}